\documentclass[12pt]{iopart}
                               
\usepackage{iopams}
\expandafter\let\csname equation*\endcsname\relax
\expandafter\let\csname endequation*\endcsname\relax
\usepackage{amsmath}
\usepackage{graphicx}
\usepackage{dcolumn}
\usepackage{bm}

\newcommand{\fat}[1]{\mathbf{#1}}
\newcommand{\fatsym}[1]{\boldsymbol{#1}}
\newcommand{\D}{\mbox{d}}

\newcommand{\te}[1]{\mathrm{#1}} 
\newcommand{\fe}{\fat{e}}

\begin{document}

\title[Dynamics of cluster formation in driven dipolar colloids dispersed on a
monolayer]{Dynamics of cluster formation in driven dipolar colloids dispersed
on a monolayer}

\author{Sebastian J\"ager, Holger Stark, and Sabine H. L. Klapp}

\address{
Institute of Theoretical Physics, Technical University Berlin, \\
Hardenbergstr.~36, 10623 Berlin, Germany 
}
\ead{jaeger@itp.tu-berlin.de}

\begin{abstract}
    We report computer simulation results on the cluster formation of dipolar
    colloidal particles driven by a rotating external field in a
    quasi-two-dimensional setup. We focus on the interplay between permanent
    dipolar and hydrodynamic interactions and its influence on the dynamic
    behavior of the particles. This includes their individual as well as their
    collective motion. To investigate these characteristics, we employ Brownian
    dynamics simulations of a finite system with and without hydrodynamic
    interactions. Our results indicate that particularly the
    translation-rotation coupling from the hydrodynamic interactions has a
    profound impact on the clustering behavior.
\end{abstract}

\maketitle

\section{Introduction}
Systems of colloidal particles can organize into a wide variety of different
structures, both under equilibrium conditions (``self-assembly'') and in
nonequilibrium. Prime examples of nonequilibrium structure formation are lane
formation of charged colloids \cite{dzubiella2002, loewen2010}, shear banding
of rod-like particles \cite{kang2006}, the coiling up of chains of magnetic
colloids \cite{casic}, metachronal waves in driven colloids \cite{wollin2011},
and the formation of colloidal caterpillars \cite{lutz2006}.

Indeed, magnetic particles form a particular subset of colloidal systems, since
they can be easily manipulated by external magnetic fields. The interplay of
particle-field interactions and the anisotropic (dipole-dipole) magnetic
interactions between the particles then leads to a wide variety of structures
such as chains \cite{martin4, martin6, martin3}, layers \cite{martin3,
martin6, leunissen, jaeger_klapp, jaeger_klapp_mhd}, and intricate
honeycomb-like structures \cite{martin5, osterman, Douglas2010}.

In this study we want to focus on the pattern formation of dipolar particles
that are driven by a rotating field. In three-dimensional systems
dipolar particles exposed to such a field tend to form layers (biaxial field)
\cite{murashov, jaeger_klapp} or membrane-like structures (triaxial field)
\cite{martin5, osterman}. In a two-dimensional geometry, in which the external
field rotates in the plane of the particles, the dipoles tend to agglomerate
into clusters. This phenomenon has been observed experimentally
\cite{weddemann2010, wittbracht2011} as well as in computer simulations that
were performed by two of the authors of this study \cite{jaeger2012}.
Clustering does not only occur for particles carrying a permanent dipole
moment, but also for particles with an induced dipole moment \cite{elsner,
tierno2007, snoswell2006}.

At suitable fields strengths and frequencies, the (permanently) dipolar
particles perform a synchronous rotation with the field. This motion gives rise
to an attractive, long-range particle interaction ($\propto -1/r^3$), which
induces a first-order phase transition between a dilute and a denser phase (in
an infinitely extended system). The observed cluster formation then corresponds
to spinodal decomposition inside of the coexistence region \cite{jaeger2012}.

It has been shown that hydrodynamic interactions play a crucial role in
numerous colloidal systems \cite{rex2008,rinn1999,dhont1996,reichert2004}. The
influence of hydrodynamic interactions on the clustering phenomenon was,
however, only shortly touched upon in \cite{jaeger2012}. Here, in this study,
we want to look into the effects that an implicit solvent has on the cluster
formation and the dynamic behavior of the particles in more detail.

The main tool we use are Brownian dynamics computer simulations with and
without hydrodynamic interactions. The particles are modeled by a
quasi-two-dimensional system of soft spheres with permanent dipole moments.
This means that the dipoles can rotate freely in all the spatial directions,
while the translational motion is restricted to a two-dimensional plane. In our
hydrodynamic simulations, we not only account for the translational
hydrodynamic couplings, but for all the couplings between the translational and
rotational motions of the particles. To understand the influence of these
different couplings is one of the goals of this study. Contrary to
\cite{jaeger2012}, we here consider a finite system size, i.e., we do not
employ periodic boundary conditions.

This paper is organized as follows: After introducing the model and the
simulation techniques, we discuss the influence of hydrodynamic interactions on
the regions of cluster formation in the field strength-frequency domain. To
understand the differences stemming from the solvent interactions, we also
consider the magnetization and synchronization behavior of the particles. In a
next step, we investigate the dynamics of the clusters and their formation.
Finally, we examine the structure of the clusters with respect to their
internal order. The paper is then closed with a brief summary and conclusions.

\section{Model and simulation methods}
In this study, we consider a quasi-two-dimensional system of dipolar colloidal
particles that are immersed in a solvent. As a model for the colloids we use a
dipolar soft sphere (DSS) potential, which is comprised of a repulsive part
$U^{\mathrm{rep}}$ and a point dipole-dipole interaction part $U^{\mathrm{D}}$:
\begin{equation}
    \label{eq:interaction}
    U^{\mathrm{DSS}}(\fat{r}_{ij}, \fatsym{\mu}_i, \fatsym{\mu}_j) =
    U^{\mathrm{rep}}(r_{ij}) + U^{\mathrm{D}}(\fat{r}_{ij}, \fatsym{\mu}_i, \fatsym{\mu}_j)
\end{equation}
In (\ref{eq:interaction}), $\fat{r}_{ij}$ is the vector between the
positions of the particles $i$ and $j$, $r_{ij}$ its absolute value, and
$\boldsymbol{\mu}_i$ is the dipole moment of the $i$th particle. The dipolar
and repulsive interaction potentials are given by
\begin{equation}
    \label{eq:dipole}
    U^{\mathrm{D}}(\fat{r}_{ij}, \fatsym{\mu}_i, \fatsym{\mu}_j)
    = - \frac{3 (\fat{r}_{ij} \cdot \boldsymbol{\mu}_i) (\fat{r}_{ij} \cdot 
    \boldsymbol{\mu}_j)}{r_{ij}^5} + \frac{\boldsymbol{\mu}_i \cdot 
    \boldsymbol{\mu}_j}{r_{ij}^3}
\end{equation}
and \cite{allentil}
\begin{equation}
    \label{eq:rep}
    U^{\mathrm{rep}}(r) = U^{\mathrm{SS}}(r) - U^{\mathrm{SS}}(r_c) 
    + (r_c - r) \frac{\D U^{\mathrm{SS}}}{\D r}(r_c),
\end{equation}
respectively. Here, $U^{\mathrm{rep}}$ is the shifted soft sphere potential,
where
\begin{equation}
    U^{\mathrm{SS}}(r) = 4 \epsilon \left( \frac{\sigma}{r_{ij}} \right)^{12}
\end{equation}
is the unshifted soft sphere (SS) potential for particles of diameter $\sigma$.

We investigate the system by making use of Brownian dynamics (BD) simulations
with and without hydrodynamic interactions (HIs). Specifically, the positions
of the particles are evolved in time via \cite{ermak1978, meriguet, dickinson}
\begin{multline}
    \label{eq:hydro_sim_T}
    \fat{r}_i(t + \Delta t) = \fat{r}_i(t) + \frac{1}{k_B T} \sum_j \fat{D}^{\te{TT}}_{ij} \fat{F}_{j} \Delta t
    \\ + \frac{1}{k_B T} \sum_j \fat{D}^{\te{TR}}_{ij} \fat{T}_j \Delta t + \fat{R}_i(\fat{D}, \Delta t)
\end{multline}   
whereas their orientations $\fat{e}_i = \fatsym{\mu}_i/\mu$ evolve according to
\begin{multline}
    \label{eq:hydro_sim_R}
    \fe_i(t + \Delta t) = \fe_i(t) + \bigg(\frac{1}{k_B T} \sum_j \fat{D}^{\te{RT}}_{ij} \fat{F}_j \Delta t
    \\ + \frac{1}{k_B T} \sum_j \fat{D}^{\te{R}}_{ij} \fat{T}_j \Delta t \bigg)
    \times \fe_i(t) + \fat{R}_i(\fat{D}, \Delta t) \times \fe_i(t).
\end{multline}   
The forces and torques in
(\ref{eq:hydro_sim_T}) and (\ref{eq:hydro_sim_R}) are given by
\begin{align}
    \fat{F}_i & = - \fatsym{\nabla}_{\fat{r}_i}
    \sum_{j \neq j} U^\mathrm{DSS} (\fat{r}_{ij}, \fatsym{\mu}_i, \fatsym{\mu}_j), \\
    \fat{T}_i & = \fat{T}_i^\mathrm{DSS} + \fat{T}_i^\mathrm{ext},
\end{align} 
where
\begin{eqnarray}
    \fat{T}_i^\mathrm{DSS} = - \fatsym{\mu}_i 
    \times \fatsym{\nabla}_{\fatsym{\mu}_i}
    \sum_{j \neq j} U^\mathrm{DSS} (\fat{r}_{ij}, \fatsym{\mu}_i, \fatsym{\mu}_j), \\
    \label{eq:Text}
    \fat{T}_i^\mathrm{ext} = \fatsym{\mu}_i \times \fat{B}^\mathrm{ext}.
\end{eqnarray} 
In (\ref{eq:Text}), $\fat{B}^\mathrm{ext}$ denotes an external field that
is homogeneous and rotates in the plane of the dipolar monolayer. Specifically,
\begin{equation}
    \fat{B}^\textrm{ext}(t) = B_0 ( \fat{e}_x \cos \omega_0 t + \fat{e}_y \sin \omega_0 t ),
\end{equation}
where $\omega_0$ is the frequency of the field and $B_0$ its
strength.

In (\ref{eq:hydro_sim_T}) and (\ref{eq:hydro_sim_R}),
$\fat{D}^\te{TT}_{ij}$, $\fat{D}^\te{TR}_{ij}$, $\fat{D}^\te{RT}_{ij}$, and
$\fat{D}^\te{RR}_{ij}$ are subtensors of $\fat{D}^\te{TT}$, $\fat{D}^\te{TR}$,
$\fat{D}^\te{RT}$, and $\fat{D}^\te{RR}$. The latter are subtensors of the
grand diffusion tensor
\begin{equation}
    \label{eq:diffusion_tensor}
    \fat{D} = 
\begin{pmatrix}
    \fat{D}^\te{TT} & \fat{D}^\te{TR} \\
   \fat{D}^\te{RT} & \fat{D}^\te{RR} 
   \end{pmatrix}
\end{equation} 
and will be specified below. The random displacements in
(\ref{eq:hydro_sim_T}) and (\ref{eq:hydro_sim_R}) behave according to
\begin{eqnarray}
    \label{eq:var}
    \langle \fat{R}_i \rangle = 0, \\
    \langle \fat{R}_i(\Delta t) \fat{R}_j(\Delta t) \rangle = 2 \fat{D}_{ij} \Delta t.
\end{eqnarray}
The actual calculation of these displacements can be done by evaluating
\begin{equation}
    \fat{R} = \sqrt{2 \Delta t} \fat{L} \cdot \fatsym{\xi},
\end{equation}
where $\fat{R}$ is the vector comprised of all the $\fat{R}_i$, $\fatsym{\xi}$
is a vector of normally distributed random numbers, and $\fat{L}$ is a lower
triangular matrix which satisfies
\begin{equation}
    \label{eq:cdecomp}
    \fat{D} = \fat{L} \cdot \fat{L}^T.
\end{equation}

In the present study, we take the HIs into account up to third order in the
inverse particle distance, which corresponds to a far field approximation. The
tensors $\fat{D}^\te{TT}$, $\fat{D}^\te{TR}$, $\fat{D}^\te{RT}$, and
$\fat{D}^\te{RR}$ are then given by \cite{meriguet, dickinson, reichert2004}
\begin{eqnarray}
    \label{eq:inn1}
    \fat{D}^\te{TT}_{ii}  = \frac{k_B T}{6 \pi \eta \sigma} \fat{I}, \\
    \label{eq:rotne2}
    \fat{D}^\te{TT}_{ij}  = \frac{k_B T}{8 \pi \eta} \frac{1}{r_{ij}}
    \left[(\mathbf{I} + \hat{\fat{r}}_{ij} \hat{\fat{r}}_{ij})
    + \frac{2 \sigma^2}{3 r^2_{ij}} (\mathbf{I} - 3 \hat{\fat{r}}_{ij} \hat{\fat{r}}_{ij})\right], \\
    \fat{D}^{\te{RT}}_{ii}  = \fat{D}^{\te{TR}}_{ii} = 0, \\
    \label{eq:trc}
    \fat{D}^{\te{TR}}_{ij}  = \fat{D}^{\te{RT \dagger}}_{ji} = \frac{k_B T}{8 \pi \eta}
    \frac{1}{r^2_{ij}} \fatsym{\epsilon} \hat{\fat{r}}_{ij}, \\
    \label{eq:inn2}
    \fat{D}^{\te{RR}}_{ii}  = \frac{k_B T}{8 \pi \eta \sigma^3} \fat{I}, \\
    \label{eq:enn2}
    \fat{D}^{\te{RR}}_{ij}  = \frac{k_B T}{16 \pi \eta} \frac{1}{r^3_{ij}}
    (3 \hat{\fat{r}}_{ij} \hat{\fat{r}}_{ij} - \fat{I}),
\end{eqnarray}
where $\eta$ is the viscosity of the solvent and $\fatsym{\epsilon}$ the
Levi-Civita density, which satifies $\fatsym{\epsilon}_{123} =
\fatsym{\epsilon}_{231} = \fatsym{\epsilon}_{312} = 1$,
$\fatsym{\epsilon}_{321} = \fatsym{\epsilon}_{213} = \fatsym{\epsilon}_{132} =
-1$, and is equal to zero for other choices of the indices. These tensors
describe the different hydrodynamic couplings, i.e., the coupling between the
translational motion of the particles (TT), between the translation and the
rotational motion (TR) and vice versa (RT), and the coupling between the
rotational motion (RR). The specific tensor given in (\ref{eq:inn1}) and
(\ref{eq:rotne2}), i.e., $\fat{D}^\te{TT}$, is the well known Rotne-Prager
tensor \cite{rotne_prager}. 

Note that by setting to zero all the hydrodynamic coupling tensors
involving different particles $i \neq j$ in (\ref{eq:inn1})-(\ref{eq:enn2}),
one arrives at the standard algorithm for BD simulations without HIs
\cite{branka1994}.

Considering the equation of motion (\ref{eq:hydro_sim_T}), we can see
that the TR coupling relates all the torques acting on the particles, i.e.,
their rotational motions, to all the translational motions of the particles.
Physically, the TR coupling describes how the translational motion of the
particles is affected by the flow fields in the solvent that are caused by the
rotations of the particles. Indeed, as we will see later, the TR/RT coupling is
particularly relevant for the overall dynamical behavior of the system. It is
therefore instructive to briefly illustrate the implications of this coupling
for a simple two-particle system (for details, see \cite{dickinson,
reichert2004}).

To this end, let us consider two particles that are located at a distance from
each other on the $x$-axis in a right-handed coordinate system. Due to the TR
coupling, anticlockwise rotation (following the application of a torque) of the
particle at the larger value of $x$ results in the other particle moving in the
negative $y$-direction. By realizing that the flow fields follow the rotation
of the particle, this process can be easily understood. On the other hand, an
anticlockwise rotation of the particle at smaller values of $x$ causes the
other particle to move into the positive $y$-direction (for a sketch, see
\cite{dickinson}).


In this study, we consider $N = 324$ particles in a simulation box that is
bounded by soft walls (cf.~\cite{grandner2008}), i.e., we do not use periodic
boundary conditions. Therefore, it is not necessary to use special techniques
(e.g., Ewald sums) to treat the long-range dipole-dipole interactions. The
forces and torques can be calculated directly via (\ref{eq:dipole}).

For convenience, we make use of the following reduced units: Field strength
$B_0^* = (\sigma^3/\epsilon)^{1/2} B_0$; frequencies of the field $\omega_0^* =
\omega_0 \sigma^2/D_0^\mathrm{T}$, where $D_0^{\te{T}}  = k_B T/3 \pi \eta
\sigma$ is the translational diffusion constant; dipole moment $\mu^* =
(\epsilon \sigma^3)^{-1/2} \mu$; time $t^* = t D_0^\mathrm{T}/\sigma^2$;
temperature $T^* = k_B T/\epsilon$; position $\fat{r}^* = \fat{r}/\sigma$. In
the following we specialize to systems at temperature $T^* = 1$ and of dipole
moment $\mu^* = 3$. This choice corresponds to a dipolar coupling strength of
$\lambda = \mu^{*2}/T^* = 9$ that is sufficiently large to enable the system to
form clusters for suitable field strengths and frequencies \cite{jaeger2012}.
The density of the particles in the simulation box is of no importance in the
investigated systems, since the particles typically agglomerate into a single
cluster.

\section{Dynamics on the particle level}
Applying a rotating field in the plane of a dipolar monolayer can cause the
dipolar particles to agglomerate into two-dimensional clusters
\cite{jaeger2012}. This agglomeration is caused by a synchronization
phenomenon: At suitable field strengths and frequencies, the particles follow
the field at the same phase difference resulting in an effective interparticle
interaction of the form \cite{martin1,jaeger2012}
\begin{equation}
    \label{eq:dipole_avg}
    U^{\mathrm{ID}} (\fat{r}_{ij})
    = - \frac{\mu^2}{2 r^3_{ij}} .
\end{equation}
Equation (\ref{eq:dipole_avg}) can be arrived at by averaging the dipole-dipole
potential over one rotational period of the field under the assumption that
$\fatsym{\mu}_i(t) = \fatsym{\mu}_j(t) = \mu_0 [ \fat{e}_x \cos (\omega_0 t +
\delta) + \fat{e}_y \sin (\omega_0 t + \delta)]$. Here, $\delta$ is some phase
difference. A further crucial assumption in the derivation of
(\ref{eq:dipole_avg}) is that the translational motion of the particles during
one rotational period of the field is negligible.

The potential (\ref{eq:dipole_avg}) is attractive, which leads to the
aforementioned cluster formation. Moreover, in an infinitely extended system,
i.e., in the thermodynamic limit, the potential (\ref{eq:dipole_avg}) gives
rise to a first order phase transition \cite{jaeger2012}.

\begin{figure}[h]
    \centering
    \includegraphics[width=80mm]{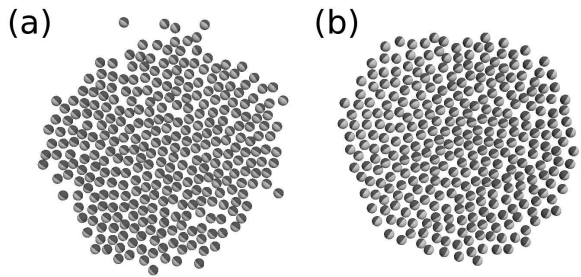}
    \caption{Snapshots of systems at $B_0^* = 50$ and $\omega_0^* = 240$ after
    clusters have formed. (a) Without and (b) with HIs.}
    \label{fig:snaps_cl}
\end{figure}
Simulation snapshots of systems after clusters have formed can be seen in
figures \ref{fig:snaps_cl}(a) and (b). The former shows the snapshot of a
system ($B_0^* = 50$, $\omega_0^*= 240$), whose particles do not interact via
HIs, while HIs are included in the system associated with the latter snapshot.
Since the formation of clusters can be observed in both the systems, we note as
a first result that cluster formation is not prevented by the presence of HIs.
As already noted in \cite{jaeger2012}, this is not a priori clear, since HIs
can induce additional motion in a nonequilibrium system and the averaged
potential (\ref{eq:dipole_avg}) is only valid as an approximation to the true
interparticle interaction if the translational motion of the particles during
one rotational period of the field is small. This is certainly the case for the
system without HIs: Inspecting the mean squared displacements (averaged over
all the particles) at $B_0^* = 50$ and $\omega_0^*= 240$, we find that the
field rotates about $30$ times before a particle transverses a distance of its
own diameter $\sigma$.

The driving frequency $\omega_0^*= 240$ chosen in figure \ref{fig:snaps_cl}
corresponds to $\omega_0 \approx 54$ MHz, if we assume the values of the
diffusion constant ($D_0^\mathrm{T} \approx 38$ $\mu$m$^2$/s) and particle size
($\sigma = 13$ nm) that are given in \cite{meriguet} for a ferrofluid.

Note that the cluster formation in infinitely extended quasi-two-dimensional
systems corresponds to spinodal decomposition within the coexistence region of
a phase transition \cite{jaeger2012}. This is not exactly true for the cluster
formation we observe in the present study. The finite systems considered here
do not undergo a phase transition. However, the used system is well suited as a
model system to investigate the influence of the HIs on the dynamic behavior of
the individual clusters.

\begin{figure}[h]
    \centering
    \includegraphics[width=80mm]{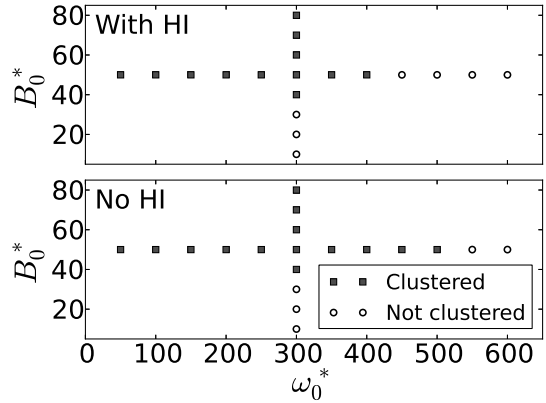}
    \caption{Cluster formation (top) with and (bottom) without HIs in the field
    strength-frequency domain. Squares/circles indicate where cluster formation
    occurs/does not occur. The systems are at temperature $T^* = 1$ and dipole
    moment $\mu^* = 3$.}
    \label{fig:hy2d_ne_diag}
\end{figure}

In the following we ask how the collective rotational behavior of the particles
changes if solvent-mediated interactions are taken into account. First,
consider figure \ref{fig:hy2d_ne_diag}, which shows whether cluster formation
occurs for selected state points in the field strength-frequency domain.
Presented are results for both the cases with and without HIs included.
Compared to the simple BD system cluster formation breaks down at smaller
frequencies $\omega_0^*$ in the hydrodynamically interacting system. In the
former, cluster formation can be observed up to $\omega_0^* \approx 450$ (at
$B_0^* = 50$) while cluster formation ceases at $\omega_0^* \approx 350$ when
HIs are present.
                                                             
\begin{figure}[h]
    \centering
    \includegraphics[width=80mm]{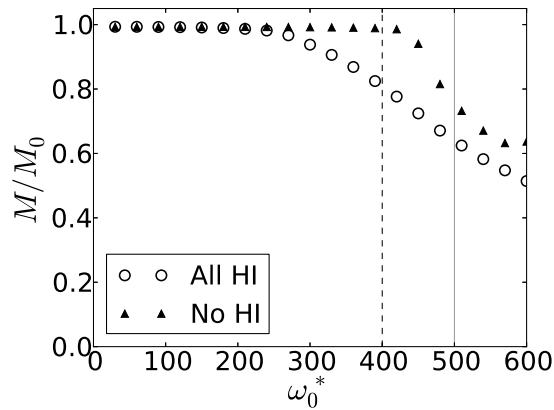}
    \caption{Magnetization normalized with respect to its saturation value over
    the driving frequency of the external field for a system with and without
    HIs. The fields used are of strength $B_0^* = 50$. The vertical line
    indicate where cluster formation ceases with (dashed) and without (solid)
    HIs.} \label{fig:magn_hy_2d}
\end{figure}

To understand the breakdown of cluster formation in more detail, we now examine
the rotational motion of the particles. As explained above, synchronized
rotation is necessary for (\ref{eq:dipole_avg}) to hold, i.e., is a
prerequisite for cluster formation. Figure \ref{fig:magn_hy_2d} shows the
absolute value of the magnetization normalized with respect to its saturation
value
\begin{equation}
    \frac{M(t)}{M_0}
    = \frac{1}{N \mu} \left< \left| \sum_{i=1}^N \fatsym{\mu}_i \right| \right>
\end{equation}
over the driving frequency $\omega_0^*$ of a system [$B_0^* = 50$,
cf.~Fig.~\ref{fig:hy2d_ne_diag}] with all the hydrodynamic couplings and
without HIs included. The magnetization indicates how aligned the particles are
in a given state, i.e., indicates if they follow the field. As can be seen, the
magnetization starts with values close to one for both the systems,
corresponding to an aligned state. At $\omega_0^* \approx 270$ the
magnetization begins to drop for the system that includes HIs. The
magnetization in the system without HIs remains at $M/M_0 \approx 1$ up to
$\omega_0^* \approx 420$.

This magnetization behavior implies that the particles in the hydrodynamically
interacting system are less aligned with each other for $\omega_0^* \gtrsim
270$. In particular, the synchronization breaks down at lower frequencies
resulting in a premature breakdown of cluster formation.

\section{Cluster dynamics}
We now turn to the dynamics of the entire cluster. In \cite{jaeger2012},
it was already shown that in systems with HIs, clusters form at a significantly
faster pace (as compared to systems without HIs). In this section, we aim to
investigate the changes induced by HIs in more detail.

\begin{figure}[h]
    \centering
    \includegraphics[width=80mm]{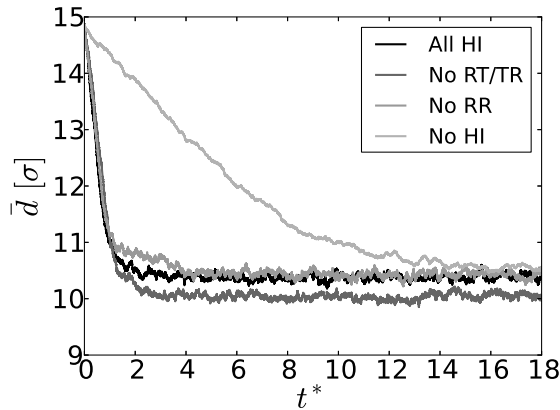}
    \caption{Mean distance between the particles over time $\bar{d}$ in a
    system ($B_0^* = 50$, $\omega_0^* = 240$) with all hydrodynamic
    interactions included (All HI), without the hydrodynamic RT/TR coupling
    (No RT/TR), without the hydrodynamic RR coupling (No RR), and without
    all HIs (No HI).}
    \label{fig:avg_dist}
\end{figure}
First, consider figure \ref{fig:avg_dist}, which shows the mean distance
\begin{equation}
    \bar{d}(t) = \frac{2}{N(N-1)} \sum_{i=1}^N \sum_{j < i} r_{ij}(t)
\end{equation}
between the particles for a system at $B_0^* = 50$ and $\omega_0^* = 240$.
These particular field parameters were chosen for three reasons: First, the
frequency is sufficiently high to ensure that the effective potential
(\ref{eq:dipole_avg}) describes the interparticle interaction well. Second, as
seen in figure \ref{fig:hy2d_ne_diag}, cluster formation occurs for both a
hydrodynamically as well as a not hydrodynamically interacting system. Third,
the magnetization of the systems with these choices of $\omega_0^*$ and $B_0^*$
is maximal and identical irrespective of the presence of HIs [cf.~figure
\ref{fig:magn_hy_2d}].

Specifically, figure \ref{fig:avg_dist} shows the evolution of $\bar{d}$ over
time for particles interacting via HIs including all the hydrodynamic
couplings, for particles lacking the hydrodynamic RT/TR coupling, particles
lacking the RR coupling, and particles not interacting via HIs at all. In all
the cases, $\bar{d}$ assumes a constant minimal value at long times. To
understand this, recall that we consider a single simulation box filled with
particles here. In an infinite system, the cluster would keep growing in time
indefinitely with a power law behavior \cite{kabrede2006,kapral2009,das2011},
since the process corresponds to spinodal decomposition \cite{jaeger2012}.
Here, however, the growth process stops once all the particles have been
incorporated into the cluster and a stationary state is reached.

In the systems that include solvent-mediated interactions (All HI, No RT/TR, No
RR), the value of $\bar{d}$ drops significantly faster than in the case without
any HIs. Consequently, the average distance between the particles decreases
faster, which means that the cluster formation process is sped up. The
acceleration is neither influenced by the lack of the RT/TR nor the RR
coupling, which implies that the TT coupling alone is responsible for this
effect. Note, however, that the lack of the presence of the RT/TR coupling
expresses itself by a different value of $\bar{d}$ at long times [see figure
\ref{fig:avg_dist}].

\begin{figure}[h]
    \centering
    \includegraphics[width=85mm]{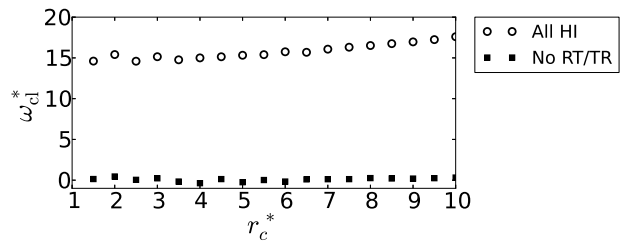}
    \caption{Mean angular frequency $\omega^*_\mathrm{cl}$ around the cluster
    center of the particles over distance from particle center $r_c^*$ for a
    system at $B_0^* = 50$ and $\omega_0^* = 240$. Shown are values for a
    system interacting via all the hydrodynamic couplings and for a system
    lacking the RT/TR coupling. }
    \label{fig:cluster_omega}
\end{figure}
The RT/TR coupling does have another interesting influence on the dynamic
behavior of the particles. In figure \ref{fig:cluster_omega}, the mean angular
frequency of the particles with respect to the cluster center over the distance
from the center is shown for a system at $B_0^* = 50$ and $\omega_0^* = 240$.
Values for the hydrodynamically interacting case with all the couplings
included as well as the case lacking the RT/TR coupling are presented. In the
system that includes the RT/TR coupling, the angular velocity of the particles
differs from zero at all the displayed distances from the cluster center
$r_c^*$. Hence, the particles perform a rotation around the cluster center in
the rotational direction of the external field. The system lacking the RT/TR
coupling does not show such a rotational behavior. As can be seen in figure
\ref{fig:cluster_omega}, the mean angular frequency of the particle around the
cluster center is essentially zero at all distances.

This collective rotation is caused by the individual, field-driven rotations of
the particles. The rotational motion of the particles creates a flow field that
induces translational motion in all the other particles [see the argument given
below (\ref{eq:inn1})-(\ref{eq:enn2})]. Therefore, the TR coupling alone is
responsible for this behavior. The RT coupling does not contribute in any way
to the cluster rotation.

As we have seen in the previous section, HIs result in a premature breakdown of
cluster formation, i.e., a breakdown at smaller driving frequencies of the
field (relative to the case without HIs). It stands to reason that the cluster
rotation induced by the HIs has a significant influence on this behavior. The
particles perform additional translational motion (around the cluster), which
makes the effective potential (\ref{eq:dipole_avg}) less accurate as a
description for the interparticle interaction at a given driving frequency. The
more the particles move during one period of the field, the less does the
effective potential capture the actual interaction between the particles.

Finally, note that the RR coupling does not seem to have any significant
influence on the dynamic behavior of the cluster. That is, it does not
contribute to the accelerated cluster formation or the cluster rotation. The
former is illustrated by figure \ref{fig:avg_dist}, which shows that the mean
distance $\bar{d}$ behaves essentially identically to the system with all the
HIs included. The fact that the cluster rotation is not influenced by the RR
coupling can be seen in figure \ref{fig:cluster_omega}. Despite the presence of
the RR coupling, the cluster does not rotate if the RT/TR coupling is absent.

\section{Internal structure of the cluster}
In a recent experimental study, Weddemann et al.~\cite{weddemann2010} showed
the existence of cluster formation in two-dimensional systems of (permanently)
dipolar particles that are driven by a rotating external field. In particular,
the authors of \cite{weddemann2010} observed the formation of hexagonally
ordered particle agglomerates in their experiments.

The general clustering phenomenon of dipolar particles exposed to rotating
fields was later identified as spinodal decomposition in a simulation study by
two of the authors of the present study \cite{jaeger2012}. However, to
reproduce clusters of hexagonal order, very low temperatures inside of the
two-phase coexistence region had to be considered in computer simulations. It
was conjectured that the hexagonally ordered clusters occur in the vapor-solid
coexistence region, i.e., at coupling strengths above the ones related to the
vapor-liquid region.

As shown in the previous section, HIs can have a significant influence on the
collective motion of the particles. Here, in this section, we want to
investigate, whether HIs preserve the internal cluster structure. Despite
experimental evidence of the hexagonal order, this fact is debatable since it
remains unclear to what degree the rotational motion of the magnetic particles
in the experimental work \cite{weddemann2010} follows the dipole moment. 

In order to gain insight into the emergent (hexagonal) structures in the
present, finite systems, we consider the bond order parameter
\begin{equation}
    \label{eq:psi6}
    \psi_6 = \frac{1}{N} \sum_{n=1}^N \frac{1}{|\mathcal{N}_n|} \left|
    \sum_{k \in \mathcal{N}_n} \exp (i 6 \pi \phi_{n k}) \right|
\end{equation}
at different dipolar coupling strengths $\lambda$. Here, $\mathcal{N}_n$ is the
set of nearest neighbors of particle $n$, which consists of particles that are closer
to particle $n$ than the distance of the first minimum in the pair correlation
function of the system. The systems considered in the following are of
sufficiently high coupling strength ensuring that cluster formation does indeed
occur.

\begin{figure}[h]
    \centering
    \includegraphics[width=80mm]{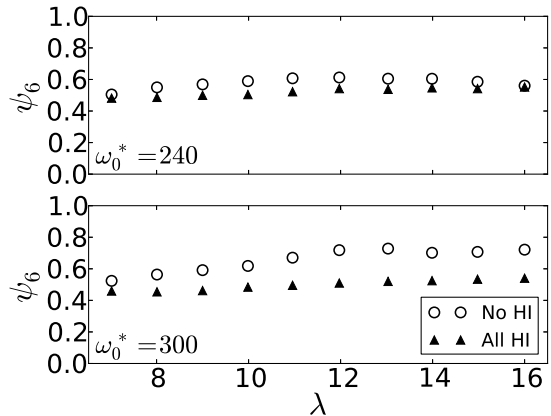}
    \caption{Bond order parameter $\psi_6$ over coupling strength $\lambda$
    for a system ($B^*_0 = 50$, $T^* = 1$)
    with and without HIs at (top) $\omega^*_0 = 240$ and (bottom) $\omega^*_0 = 300$.}
    \label{fig:psi_hy}
\end{figure}
In figure \ref{fig:psi_hy}, $\psi_6$ as function of the dipolar coupling
strength $\lambda$ for systems ($B_0^* = 50$) with and without HIs at
$\omega_0^* = 240$ and $300$ is shown.
We note that $\psi_6$ is essentially independent of time in the stationary
situation, where all the particles are part of the cluster. At $\omega_0^* =
300$, the bond order parameter increases with $\lambda$ for both the
hydrodynamically as well as the not hydrodynamically interacting system.
If HIs are not present, an increase from $\psi_6 \approx 0.52$ at $\lambda =
7.02$ to $\psi_6 \approx 0.72$ at $\lambda = 16$ can be observed. This increase
qualitatively agrees with the one observed in the Langevin dynamics simulations
in \cite{jaeger2012}. In the system with HIs included, the bond order parameter
$\psi_6$ increases considerably less with increasing coupling strength.
However, there is still significant order in the system.

In the system with $\omega_0^* = 240$ and HIs included, $\psi_6$ behaves
similarly to the case with $\omega_0^* = 300$ and HIs. If these interactions
are not taken into account, however, less order than in the $\omega_0^* = 300$
system can be observed. The smaller frequency allows for more translational
motion during one rotational period of the field, resulting in more spatial
inhomogeneity.

\begin{figure}[h]
    \centering
    \includegraphics[width=80mm]{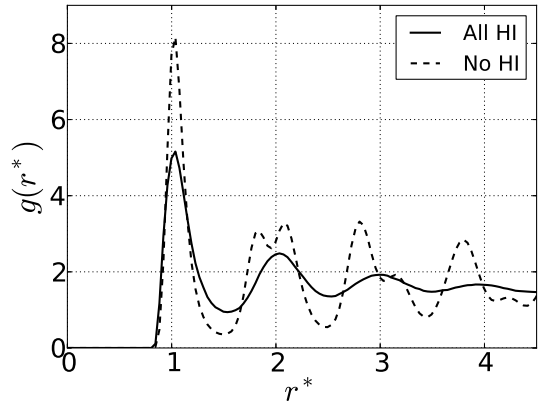}
    \caption{Pair correlation functions of systems at $B_0^* = 50$ and
    $\omega_0^* = 300$ with and without HIs included.}
    \label{fig:gr_hy}
\end{figure}

Now consider figure \ref{fig:gr_hy}, which shows the pair correlation functions
of a system ($B_0^* = 50$, $\omega_0^* = 300$) with and without HIs included.
The pair correlation function of the not hydrodynamically interacting system
shows a double peaked maximum after the first minimum, which is typical for a
hexagonally ordered system. The hydrodynamically interacting system, on the
other hand, does not have this feature. Hence, the particles tend to have six
angularly equally distributed neighbors as shown by the value of $\psi_6$
[cf.~figure \ref{fig:psi_hy}], but seem to lack the long-range positional order
of a hexagonally structured system. Further, the extrema are much more
pronounced in the system without HIs, indicating a more ordered state.

In conclusion, in both systems in figure \ref{fig:psi_hy}, the HIs tend to
weaken the hexagonal (or hexatic) order present in the system. The lower value
of $\psi_6$ in systems with HIs can be explained by the collective cluster
rotation induced by the hydrodynamic TR coupling. The particles rotate around
the cluster center and do not stay at fixed lattice sites. This behavior
results in a reduction of the bond order parameter and the hexagonal structure
in the system.

\section{Conclusions}
In this study, we have investigated the influence of the solvent on the dynamic
behavior of rotationally driven dipolar particles.

The cluster dynamics and formation is influenced in three major ways by HIs.
First, HIs accelerate the cluster formation process. While the particle
approach each other, flow fields are created that drag other particles along.
Second, we have shown HIs to induce a collective cluster rotation. Without HIs
no such rotation can be observed. The driven rotating particles pull the
solvent with them in their rotational motion, which results in the
translational motion of the particles around the cluster center. As a last
major point, we have shown cluster formation to cease at lower driving
frequencies of the field. We attribute this to an earlier breakdown of
synchronization if HIs are present and an increase in translational motion due
to the collective cluster rotation. 

Moreover, we have studied the influence of HIs on the internal structure of the
clusters. Our results indicate that HIs tend to decrease the hexagonal order in
the systems.

Recently, one of the authors of this study has conducted an investigation of
the structure of a closely related system \cite{starkpriv}. The system
consisted of particles that only interact via HIs, are confined to a monolayer,
and have fixed angular velocities. It was found that hexagonal particle
agglomerations rotate, with the hexagonal order melting and recrystallizing
periodically. In the system focused on in the present study, no such phenomenon
could be observed. We attribute this to the presence of dipolar interactions in
our system and a lack of Brownian motion in \cite{starkpriv}.

Another study has recently investigated the interplay of confinements and HIs
\cite{goetze2011}. It was shown that particle rotations can be utilized to
create directed translational motion in colloids in specific geometries. This
result suggests that it might be interesting to examine the effects of
different confinements on the dipolar systems considered here.

In conclusion, we have shown that HIs have a considerable influence on the
formation and the dynamics of clusters of driven dipolar colloidal particles in
a two-dimensional geometry. Usually, HIs seem to affect colloidal
nonequilibrium clustering phenomena less significantly than in the present
system. As an example, consider the process of colloidal gelation in
two-dimensional Lennard-Jones systems \cite{yamamoto2008}. In contrast to our
system, the agglomeration of the particles in \cite{yamamoto2008} is only
marginally influenced by the HIs. In general, however, HIs can significantly
alter nonequilibrium processes. For instance, HIs can enhance ratchet effects
\cite{grimm2011,malgaretti2012} or synchronize the motion of eukaryotic
\cite{goldstein2009} or bacterial \cite{reichert2005} flagella.

\ack
We gratefully acknowledge financial support from the DFG within the research
training group RTG 1558 {\em Nonequilibrium Collective Dynamics in Condensed
Matter and Biological Systems}, project B1.

\section*{References}

\end{document}